10 October 2023

# The Impact of an Innovative Education and Outreach Project by a Physics Experiment

**R. MICHAEL BARNETT**
*Lawrence Berkeley National Laboratory, Berkeley CA 94720*
**K. ERIK JOHANSSON**
*Professor Emeritus*
*Stockholm University, Stockholm, Sweden*

## ABSTRACT

When the education and outreach project of ATLAS – a leading experiment at the Large Hadron Collider – was initiated in 1995, we wanted to share the drama and excitement in the exploration and discovery of new science. The goal was to make these fascinating stories understandable and available to everybody and inspire the next generation of scientists and engineers. The education and outreach material included classroom activities, an extensive website, videos, a YouTube channel, social media, a planetarium show, and materials related to search for and discovery of the Higgs Boson.

## Introduction

Physicists at the Large Hadron Collider (LHC) [1] in Geneva are excited when they discuss the secrets of the universe that are being searched for. When the education and outreach project of the ATLAS Experiment [2] was initiated, we wanted to share the drama and excitement in the exploration and discovery of new science, engage students in science and inspire the next generation of scientists and engineers.

Starting around 1995, physicists in the ATLAS Experiment at the LHC decided that we wanted an extensive and systematic program of education and outreach that would have true impact, and which we were prepared to devote real effort to achieve. The aim was to reach and enthuse the general public and students with the physics of ATLAS and how we learn about physics through experimentation. With physicists from thirty-eight nations, ATLAS has an unparalleled opportunity to share the excitement of new physics with a worldwide audience.

Our goal was to make these fascinating stories understandable and available to everybody. Many people made outstanding contributions to the work of the ATLAS Education and Outreach group during that period (see the Acknowledgements). And this work continues up to this day. Our audiences include the general public teachers and students, visitors to CERN (more than 60,000 as of 2011), scientific magazines, the news media (both print and television/radio) and bloggers.  This broad audience is excited and motivated by learning about the amazing developments in contemporary physics.

# Impact of Education and Outreach by ATLAS Experiment

The ATLAS Experiment is gigantic both in the size of the apparatus (almost as big as the Notre Dame Cathedral in Paris) and the number of collaborators (more than 3000). The physicists constructing and running one of the largest physics experiments ever constructed were located at over 180 universities and laboratories around the globe (over 40 countries). Many of them took part in this outreach endeavour and contributed to the global impact. This article describes the period 1996-2012, when ATLAS outreach began, a work that is still ongoing.

In order to describe and explain the fundamental processes in particle physics, we wanted to use the many techniques available to produce interesting material for education and outreach purposes and highlight some of the most exciting explorations of the ATLAS experiment. This included giving students the possibility to explore the particles and processes using real particle collision data from ATLAS. The impact of this effort is described in the following 14 short sections.

## 1) Website

The ATLAS website is the portal to most of our education and outreach information for students, teachers, and the public. This website had almost seven million hits in 2012 (an average of 132,000 hits per week). Through 2012, there were a dramatic 26 million hits on atlas.ch website (which has been renamed as atlas.cern). See the distribution in Fig. 1. The site had 700 webpages at that time.

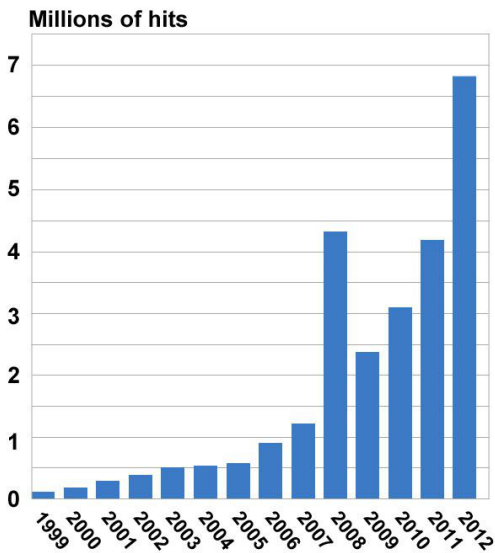

Fig. 1: Millions of hits on the ATLAS website in 1999-2012.

There were specialized webpages for the news media and for students and teachers. The press page contained in compact form interesting images and video accompanying the physics news releases and developments. The student/teacher pages had everything from fun pages to the student event analysis sites.

The Discovery Quest introduced readers to the potential discoveries that motivate the experiment,

covering: dark matter, supersymmetry, the origin of mass, extra dimensions, antimatter, new forces, the unknown, as well as the Standard Model. The ATLAS eTours had sections on physics, experiment, and accelerator. Together they encompassed 60 pages. The information on social media answering basic questions of physics, the nature of the collaboration and the people in ATLAS have constantly grown over time. The ATLAS Run and LHC run section provided the latest information about the accelerator and particle collision live events, as well as access to relevant photos, videos and DVDs.

From this single website portal, one could discover an amazing world of information about ATLAS and about physics.

## 2) YouTube Channel

Another mode of communicating science was the ATLAS YouTube channel at YouTube.com/TheATLASExperiment. It had 38 videos (of 86 videos produced by ATLAS Outreach). The top one had 134,000 viewings. In total, there were 840,000 viewings as of 2012. Eleven clips were rated as 5 stars (highest). Some of the resulting comments from viewers were:
- Very awesome channel… Cheers!!!
- Wow I love your channel!! It's awesome!
- Great channel.
- Wow!!!

## 3) Videos

One of the formats for communicating science produced by the ATLAS education and outreach group were many videos about the physics and the construction, operation, and experimental techniques. We even mounted a camera on a giant toroid magnet as it was lowered into the cavern far underground. Two early animations (in Star Wars style) were.
ATLAS – Episode 1 – A New Hope
ATLAS = Episode 2 – The Particle Strike Back

Among the many videos was the so-called "ATLAS Experiment Movie". This movie won Gold Medals at four international film festivals (produced in ten languages). It gives a glimpse behind the scenes of building the ATLAS detector. This film asks:
- Why are so many physicists anxious to build this apparatus?

Will they be able to answer fundamental questions such as:
- Where does mass come from?
- Why does the Universe have so little antimatter?
- Are there extra dimensions of space that are hidden from our view?
- Is there an underlying theory to find?

Major surprises are likely in this unknown part of physics.

## 4) Exploring particle collision events: A Hands-On Experience

By studying particle collisions at very high energies, the ATLAS experiment is exploring how particles are produced and how the fundamental processes affect the inner structure of matter. One of the main aims of ATLAS has been to shed light on what mass is. This has for a long time been a mystery. The prevailing theory at the time (the Brout-Englert-Higgs mechanism, sometimes just called the Higgs mechanism) described a mechanism that gives rise to mass. This mechanism predicted the existence of a particle, the Higgs particle. Finding and observing the Higgs particle was the confirmation of the mechanism that gives rise to mass.

A lot of the material we produced in the outreach group aimed at describing the detector and how it is used to explore the normally hidden processes in microcosm. But particularly for teachers and students we wanted to go further and explain the dynamics of the constituents of matter, and make students involved in the exploration of particle collisions.

When two protons collide at high energy, it is inside the protons that the real action is taking place. It is the constituents of the protons, which are the quarks and gluons, that interact. Our aim was to intrigue and inform students and teachers with these fundamental processes in the interior of matter [3]. One of the most important ingredients in this education effort was actual particle collisions that students could study.

Particle collisions are dramatic, Fig. 2. A large amount of energy is concentrated on the two particles colliding in the middle of the ATLAS experiment. In the very early universe, when it was only a fraction of a nanosecond old, these particle collisions dominated. In images and films, we have tried to exhibited the intensity of these dramatic particle collisions. In ATLAS Episode 2 (https://videos.cern.ch/record/1096390) the particle collision is accompanied by dramatic music similar to that of Star Wars.

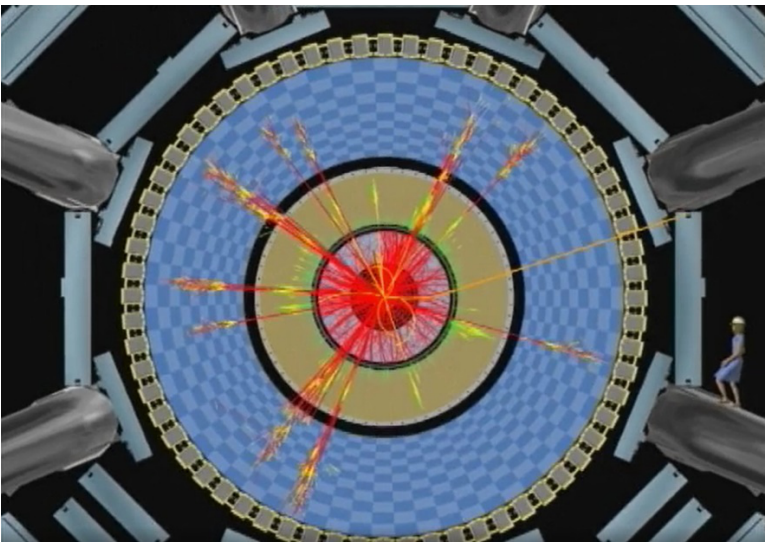

Fig 2: A particle collision occurring in the detector. The woman is shown to indicate the scale.

### 4.1 Short-lived particles in ATLAS

In the high energy particle collisions a large number of particles are produced. Among this multitude of particles, some particles originate from the decay of particles produced in the collision that leave no tracks. To reconstruct these invisible particles is something a student can manage to do with some help from a physicist.

The decay point of particles with lifetimes around $10^{-10}$ seconds can be reconstructed with the precise information of the detector data. The decay can be visualised, and the event can be analysed with reconstruction techniques that the students can learn with some guidance, see Fig. 3. Quantities like electric charge, mass and lifetime can be determined [4].

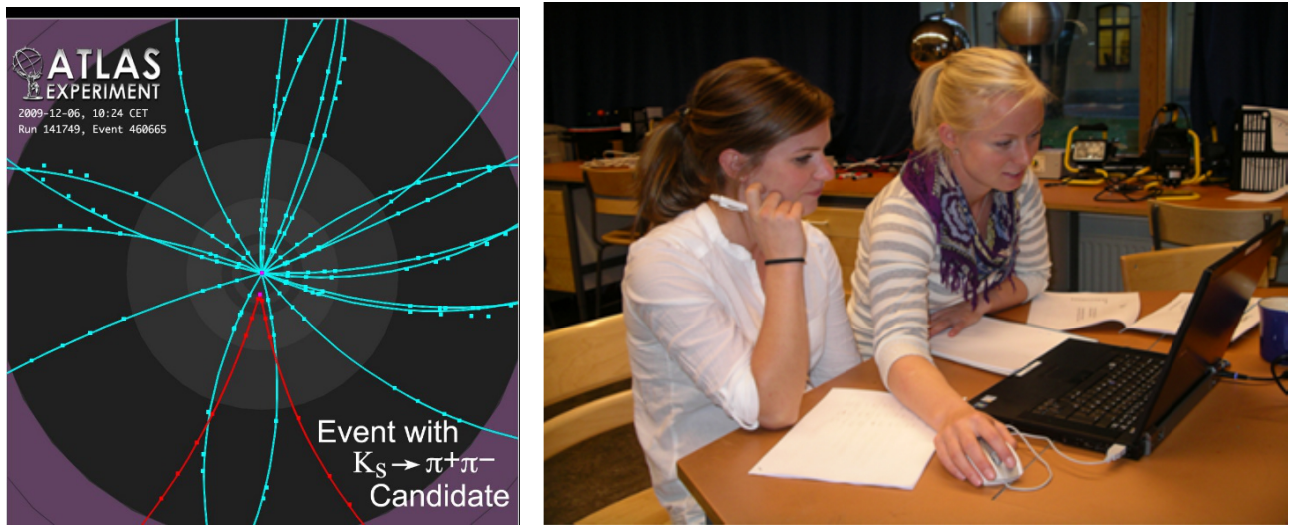

Fig. 3: a) A neutral short-lived particle decaying into a positive particle and a negative particle in the Inner Detector of ATLAS (red particle tracks). b) Students in Stockholm House of Science exploring the decay of short-lived particles produced in ATLAS.

A similar technique can be used to search for and reconstruct the elusive Higgs particle. A difference is the extremely short lifetime of the Higgs particle, which is so short that it will decay practically immediately, and the decay point will be invisible.

### 4.2 The Higgs particle

One of the main aims of the ATLAS experiment was to shed light on the origin of mass and to ultimately find the Higgs particle. Many of the products of the education and outreach group, like the artistic particle collision images, the physics brochure as well as other education material, focused on the search for the Higgs particle.

Similar techniques that are used to reconstruct and explore particles with lifetimes around $10^{-10}$ seconds are used to reconstruct the extremely short-lived Higgs particle. An animation of the dynamics involved in the production and decay of the Higgs particle can be seen in [5]. The people visiting ATLAS and the students taking part in the Masterclasses get a chance to learn more about the Higgs particle and the dynamics behind the origin of mass.

After a long search, the discovery of the elusive Higgs particle was announced in 2012 by the ATLAS and CMS experiments [6]. The discovery resulted in an increased interest in ATLAS, which can be seen in the huge number of website hits that year. Today the ATLAS mural is adorned with a collision event in which a Higgs particle has been produced.

Fig. 4 shows a particle collision in which a Higgs particle has been produced.

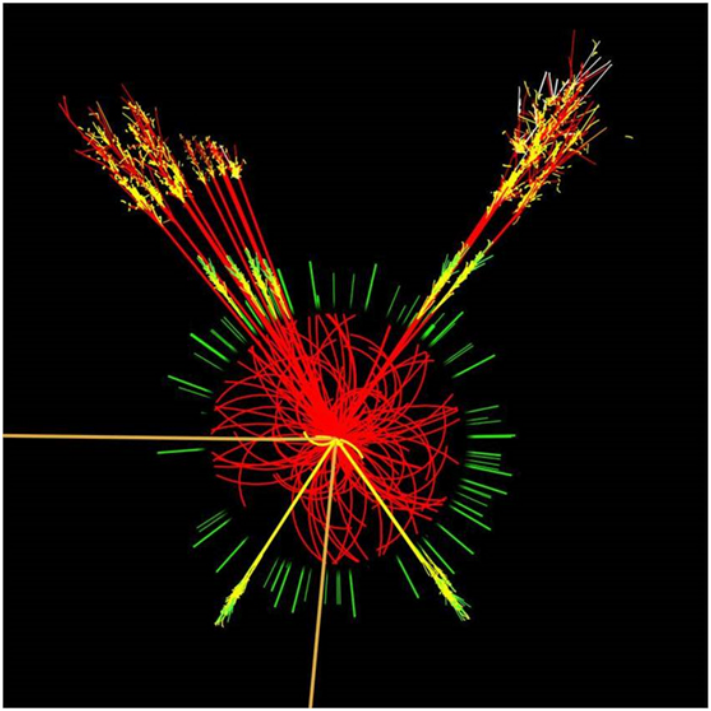

Fig. 4: An ATLAS collision event in which a Higgs boson was produced and quickly decayed to the particles observed in the detector.

### 4.3  Learning with the ATLAS Experiment

Our aim was to present contemporary physics at school to enthuse students and teachers about today's frontline physics and support the members of the collaboration in their education and outreach activities. In addition to informational material like films, animations and brochures, several analysis programs were created for high school students to make it possible for them to analyze samples of actual ATLAS events, which were tested at worldwide masterclasses (ATLAS Resources [7]).

With these tools, students can explore the same particle collisions that thousands of scientists around the world use in their research. Experimental data from the ATLAS Experiment have also been made available for students and teachers participating in the International Masterclasses in particle physics. The International Masterclasses [8] now attract around 13,000 college students from around 60 countries every year.

For a long time, the ATLAS education and outreach project has collaborated with the International Masterclasses [8] and the International Particle Physics Outreach Group [9] to make it possible for students to take part in the exploration of particle collision data. This education collaboration is still very much active and continues to expand.

### 5) Pop-up Book, a Special Event, and News Coverage

We worked with author Emma Sanders and artist Anton Radevsky to create a spectacular pop-up book for the public.  It was introduced at a well-attended special event that was held at the New York Academy of Science in the World Trade Center building that survived the 2001 attack.  It featured actor Alan Alda as host, and Lisa Randall, Michael Tuts, Emma Sanders, and Anton Radevsky as speakers.   See Fig. 5.

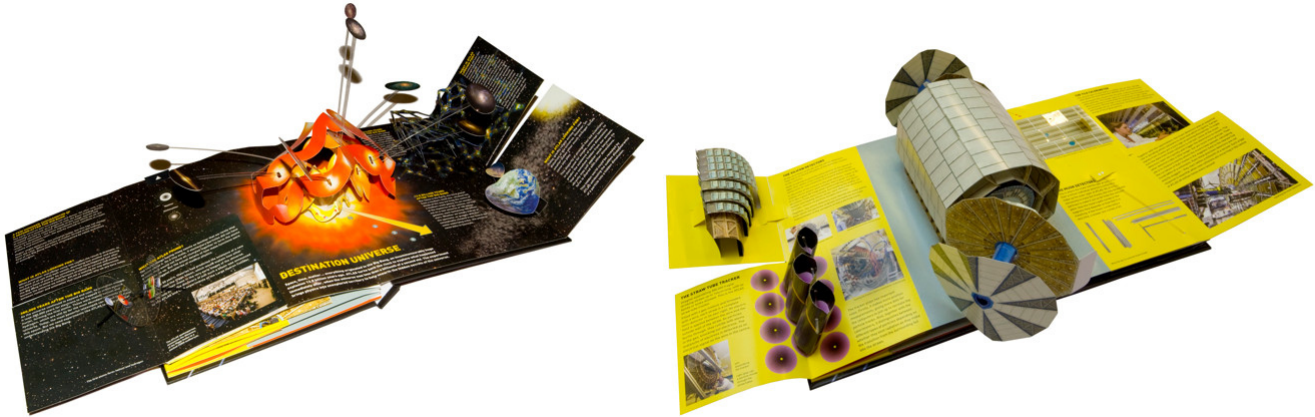

Fig. 5:  Two pages from Pop-up book

There was news coverage of pop-up book in Los Angeles Times, Discovery Channel, PC World, Gizmodo and more.  Gizmodo's headline said**:**
> **"This is Simply the Coolest Pop-up Book We've Seen."**

### 6) Giant Mural and News Coverage

We commissioned mural artist Josef Kristofoletti to produce a three-story mural (2010) of the ATLAS detector near the control room.  The ATLAS mural has attracted huge attention, see e.g. Fig. 6.

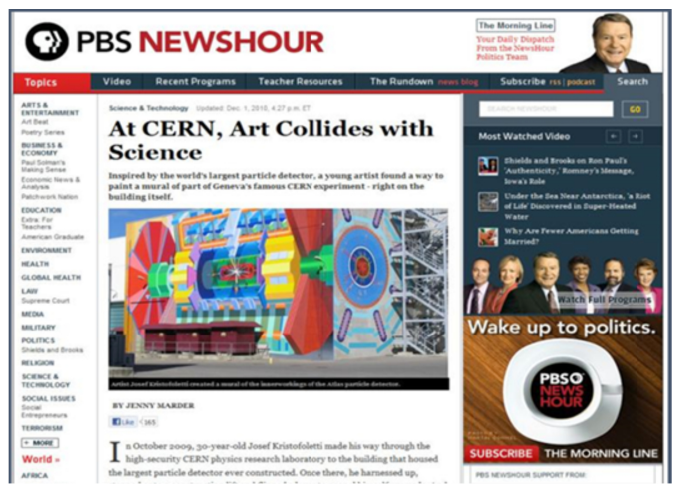

Fig. 6: ATLAS mural in PBS Newshour (also New York Times, the Guardian, and many other news media).

## 7) Artistic Views of ATLAS Collision Events

We have created multiple types of views of ATLAS collision events, such as those below. The aim is to attract and enthuse our audiences with the nature of our data.  See Fig. 7.

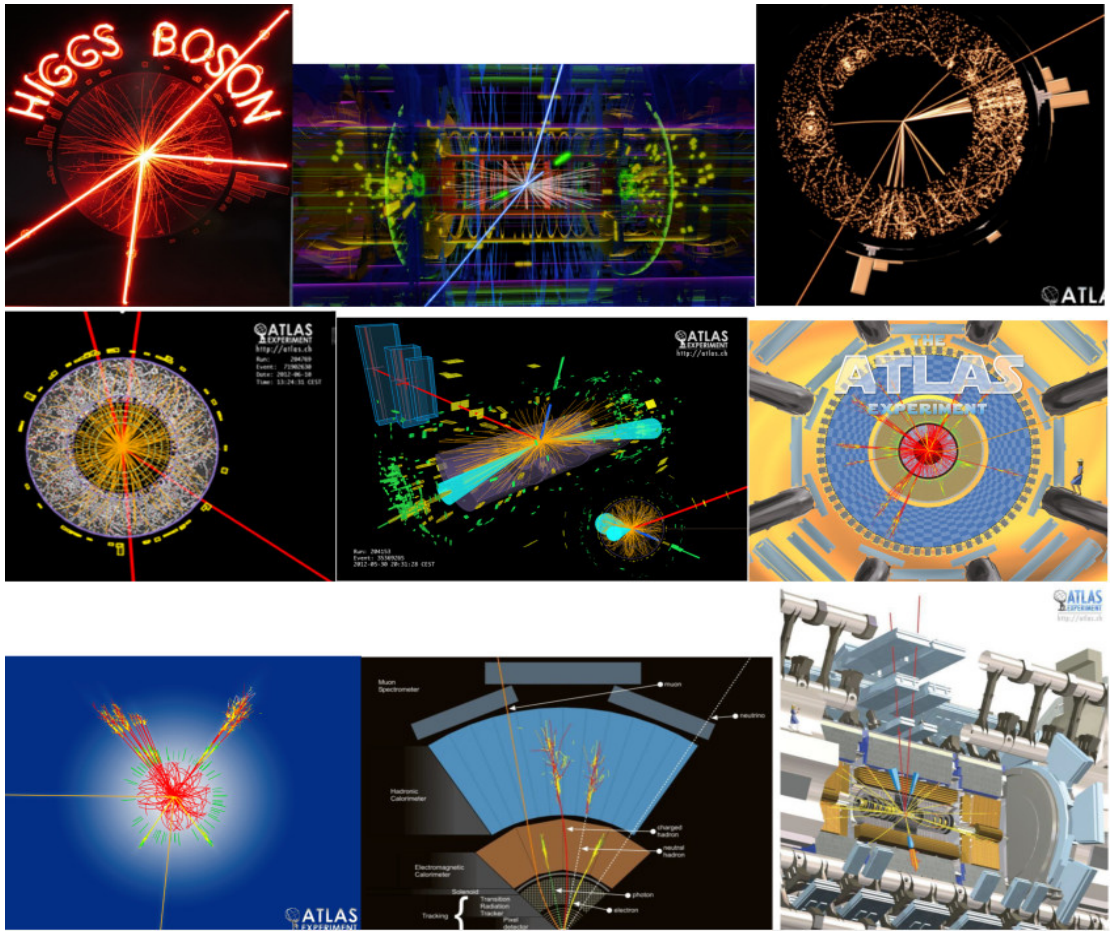

Fig. 7:  Artistic Views of ATLAS Collision Events.

## 8) A Multitude of Printed Materials

ATLAS created many printed materials in addition to the Pop-up book, including brochures: General, Physics, Tech Transfer, Technical Challenges, Extra Dimensions, Computing, Fact Sheets**,** and press kit.  Our general brochure changed the appearance of how brochures look. It was a big "hit" and reprinted again and again.  Se Fig. 8

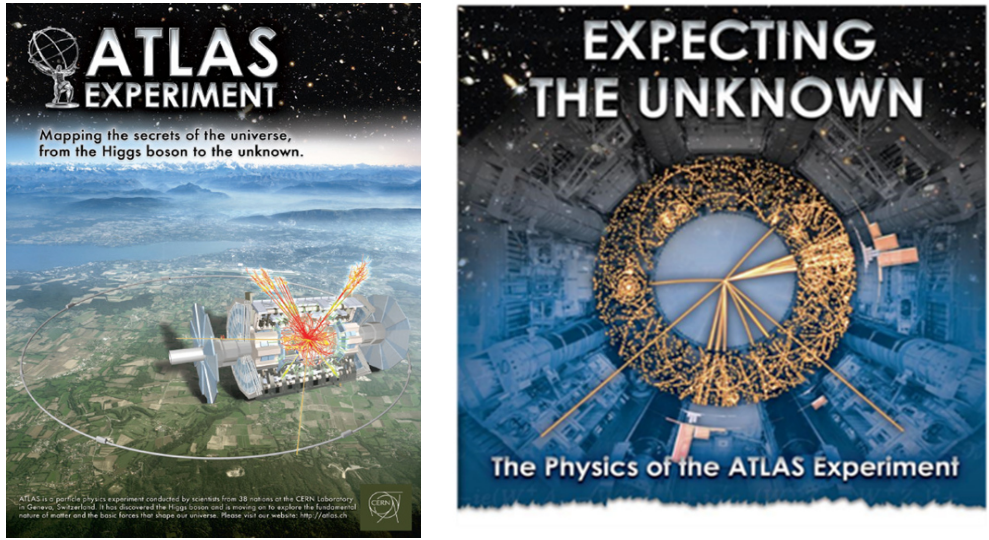

Fig. 8: The cover pages of an ATLAS general brochure and a brochure about our physics.

### 9) Photoshop Contest with Many Entries

ATLAS Outreach organized a Photoshop contest using a classic photo of ATLAS under construction. It produced many amazing images, such as these in Fig. 9:

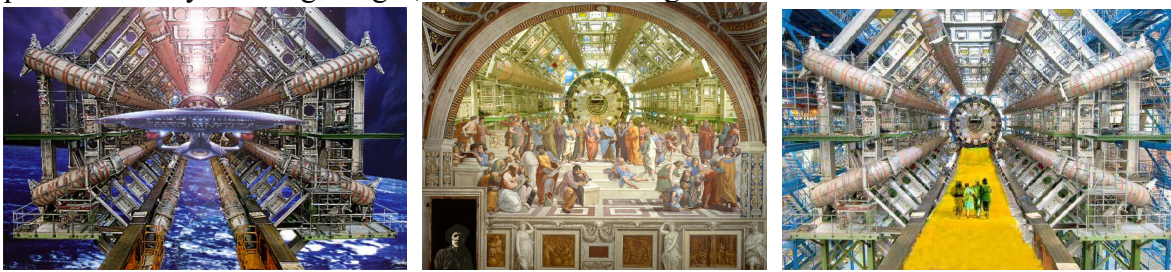

Fig. 9: Three of the entries received for our ATLAS Photoshop contest.

### 10) LEGO Model of the ATLAS Detector

ATLAS Outreach developed a LEGO Model of the ATLAS experiment with almost 10,000 pieces. 68 institutions worldwide have constructed this model for display.

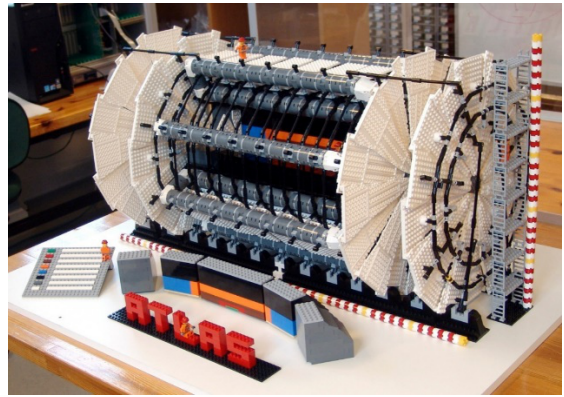

Fig. 10: One of the many constructions of the LEGO model of the ATLAS detector.

### 11) Planetarium Show

A planetarium show developed and produced by ATLAS members has had massive impact.
It is in more than 750 planetariums in 74 countries and has been translated into 27 languages.
It has been seen by more than 2 million people. At an International Full Dome Festival, the show won an award for Outstanding and Innovative Production. See the show's website at:
https://phantomoftheuniverse.org/

The planetarium show "Phantom of the Universe" explores the mysterious dark matter, which is believed to be the dominating component of matter in the Universe. This unknown matter could be due to particles searched for by ATLAS and other experiments at the Large Hadron Collider.

The show covers dark matter from the Big Bang to galaxies to underground searches to the LHC and ATLAS. It is a state-of-the-art show with an Academy Award winning narrator for the English version, Tilda Swinton. Since the producers had no experience with planetarium shows, we choose real professionals. Among them were a Hollywood scriptwriter and a Hollywood producer. Sound was handled by Skywalker Sound, which has won Academy Awards for their work. We consulted with experts from seven planetariums.

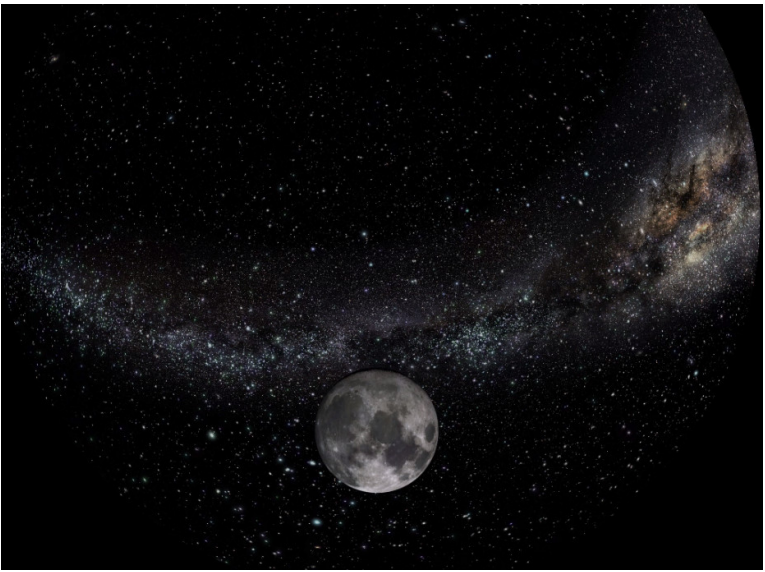

Fig. 11: A scene from the planetarium show, Phantom of the Universe.

## 12) The ATLAS Visitor Centre at CERN

The ATLAS Visitor Center is an arena for education and outreach, using informational material described in this article. It is situated close to the ATLAS mural and next to the ATLAS control room. The ATLAS mural is a good meeting point for visits to ATLAS (Fig. 12). About 30,000 people a year pass through the visitor centre.

The particle collisions take place 100 m under the visitor's feet in the underground cavern of the ATLAS detector. The visits to the ATLAS Visitor Centre are often guided by research students, who also describe their own role in the experiment. The guides often get competition from the live view of the ATLAS control room with the physicists busy verifying that all the detector components are functioning correctly.

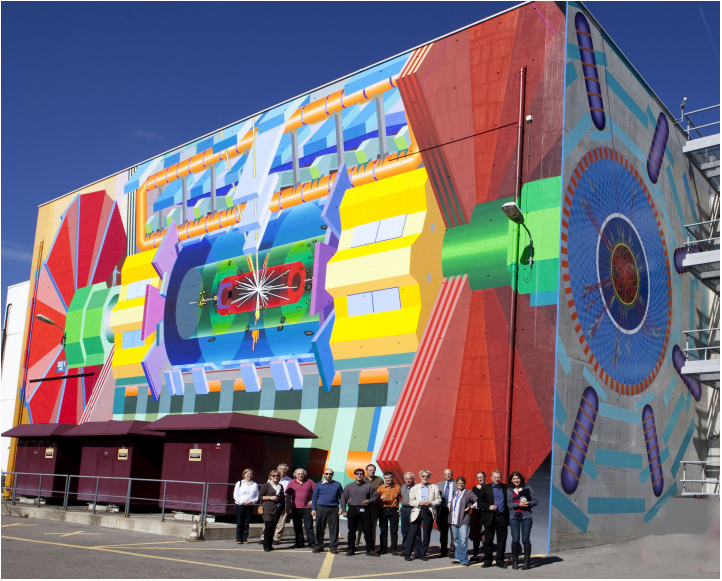

Fig. 12: The three-story tall mural of the ATLAS detector near the ATLAS control room.

## 13) Cultural Impact

Our efforts also had cultural impact.

### 13.1 Angels & Demons movie

A major Hollywood movie, Angels & Demons, with Tom Hanks had five minutes about the Large Hadron Collider, and much was filmed inside the ATLAS cavern. ATLAS Outreach worked with Sony Pictures to create an extra segment about the ATLAS experiment on the DVD of the movie. We travelled to their studios near Hollywood to organize and plan the DVD extra segment. We also created a website for the physics of Angels & Demons, an antimatter booklet, and a PowerPoint template for talks centered around the movie and ATLAS. See. Fig. 13.

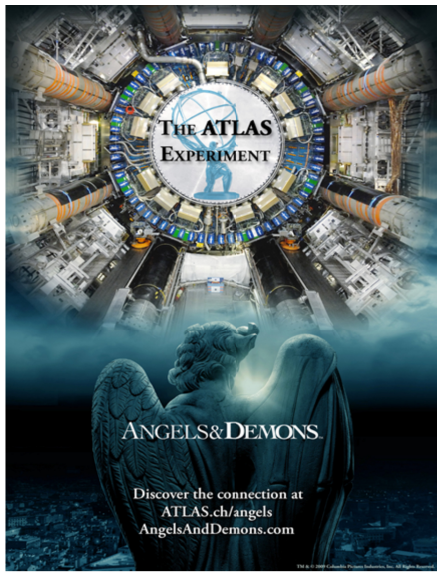
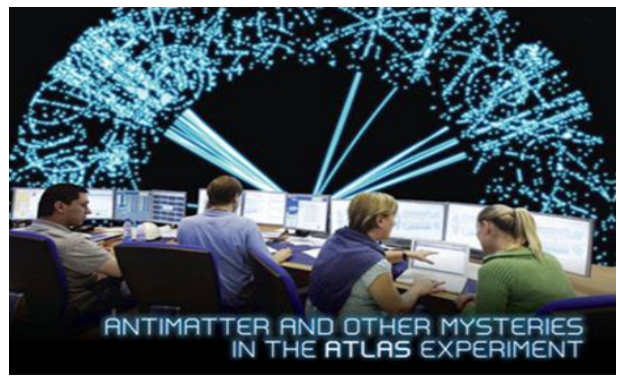

Fig. 13: a) The ATLAS website for Angels & Demons related materials, and b) The Antimatter Booklet.

## 13.2 Berlioz opera
Our efforts also had cultural impact in an opera: The backdrop of a scene in the opera "Les Troyens" by Hector Berlioz was an adaptation of a photo of the ATLAS detector under construction. This opera was shown in Valencia, St. Petersburg, and Warsaw.

## 13.3 Muppets movie
In a 2011 movie partly filmed in the ATLAS cavern, the Muppets go to CERN In the movie, Kermit the Frog sets to reunite his gang of muppets, and his journey takes him all the way to CERN.

## 13.4 The Big Bang Theory show
The Big Bang Theory comedy show, where two physicists explain the mysteries of physics, was aired on CBS from 2007 to 2019. It received several Emmy awards. Fig 14 shows the ATLAS physics poster in the TV show.

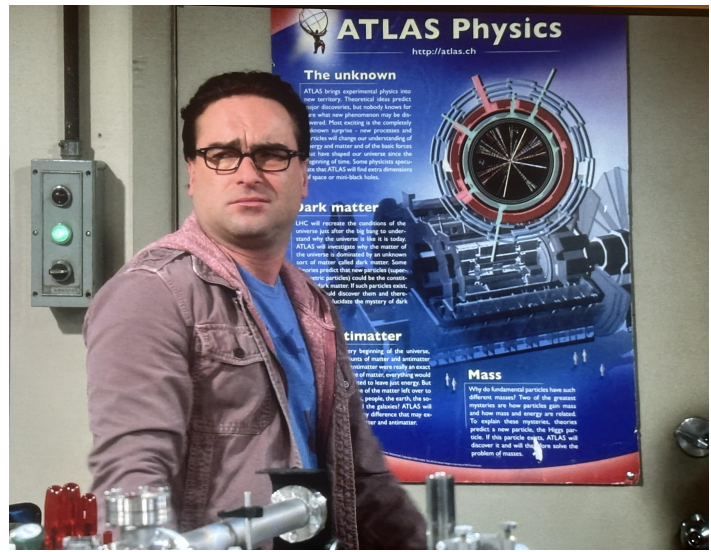

Figure 14: One of the actors and the ATLAS physics poster in the Big Bang Theory show.

## 13.5 PhD comics
The famed author of PhD comics, Jorge Cham, made some comic strips about ATLAS describing a young physicist's role in this army of scientists and enormous amount of data.

## 13.6 Scientific American article on ATLAS Art:
"19 Ways that Art and the LHC Open a Portal to Physics"
The 19 images of the Scientific American art article [10] show the different artistic expressions used to visualize the dynamic nature of exploring particle collisions and the secrets of microcosm.

## 13.7 Variety of media impact
Injecting excitement with an aesthetic touch in the presentation of the ATLAS Experiment seems to have caught the attention of a variety of media and also inspired a variety of formats. An opera, a Hollywood movie, a Muppet movie, a TV comedy show and a comics series demonstrate that ATLAS has made it to arenas where physics is not commonly found.

## 14)  The ATLAS education and outreach project

### 14.1 Exploring new routes for outreach in ATLAS
When we started the outreach project in the ATLAS collaboration, the aim was to provide information about particle physics and the ATLAS experiment to a large international community using the most recent technology. We designed new brochures with a new look using interesting photos of the detector taken by an experienced photographer. These photos quickly spread, and were seen to appear in newspapers and magazines all over the world, even in the decor of an opera.  Equally important were the animations of particle interactions, which showed the numerous particles emerging from the collision going through the many detectors of ATLAS. These particle animations were also an important component in the films produced.

The outreach activities were very well received within the collaboration, and the strong support from the collaboration was crucial for its success.  Many took active part in the project, and the collaboration agreed that the education and outreach project should be supported.

### 14.2 Inspiring and working with other experiments
The success of the education and outreach project in ATLAS also inspired other experiments. After some time all the LHC community was interested in learning from each other, and a dynamic outreach community was formed with CERN and many of the experiments at the LHC accelerator

### 14.3 Collaborating with International Masterclasses and IPPOG
The aim of the International Masterclasses and the International Particle Physics Outreach Group (IPPOG) to make it possible for high school students all over the world and the interested general audience to learn about the processes inside matter, coincided with that of the ATLAS education and outreach project.  This resulted in a long and close collaboration between ATLAS, the International Masterclasses and IPPOG to make it possible for students to take part in the hands-on exploration of particle collision data. This collaboration is still going on and expanding.

## 15) Summarizing the Innovative Achievements of Education and Outreach in the ATLAS Experiment

**Website and YouTube site**
The public website was the hub for most of the information about the ATLAS experiment and the physics that ATLAS was exploring. There were typically 3 or 4 million hits per year, but with the discovery of the Higgs boson in 2012, it jumped to 7 million, showing the importance of the website to students and the public. Together with the ATLAS YouTube site, it formed a popular and important site for information about physics.
**Animated particle collisions and artistic views of particle collisions**
Probably one of the most successful projects were the animated particle collisions and the artistic views of particle collisions. They were frequently used by the ATLAS community in interactions with students and the public, and they showed up in many magazines and TV programs. Other experiments soon had similar animation projects.
**Exploring particle collisions**
Animations of particle collisions was an attractive way of showing what happened when particles collide. For students a more hands-on project to explore short-lived particles having a lifetime of $10^{-10}$ seconds was developed. This was intended to be the first step to explore the decay of events with a Higgs particle with very much shorter lifetime. For this exercise the analysis programs Minerva (Rutherford Appleton Laboratory and Birmingham University) and Hypatia (University of Athens) were used. The analysis programs Minerva and Hypatia are still in use in International Masterclasses [8] for the analysis of the W and Z particles – the mediators of the weak force. In Masterclasses, the students get the guidance needed to explore real particle collisions.

**Lego model and Mural**
The production of a Lego model and the painting of the ATLAS mural are two very different projects showing the ATLAS detector. 68 institutions worldwide have constructed the Lego model for display. Tens of thousands of visitors have passed the large mural on their way to the ATLAS visitor centre. Both these projects are rather unique for ATLAS.

**Planetarium show**
The successful planetarium show is rather unique for ATLAS. Production involved many people with different skills. It has been seen by over two million people in 750 planetariums in 74 countries and is in 27 languages.

**Subjects that seemed to particularly interest students and the general public:**
- New discoveries and exploring the unknown.
- Learning about and exploring particle collisions.
- Visiting the experimental area.
- Contact with physicists.

## Conclusions

During 1996-2012 the education and outreach project in ATLAS was formally established and soon became an integrated part of the experiment. The ATLAS education and outreach program has continued to this day, and has made many major advancements, including those in social media, exhibitions, virtual visits and classrooms visits. This process has continued to spread, resulting in education and outreach activities having become a natural part of international physics projects.

- Today all major physics experiments have an outreach project integrated in the collaboration - a revolution in frontline physics education and outreach, which we are happy to have been part of.
- Future experiments now often have an ambitious education and outreach project. Many research councils now require applicants to have an education/outreach project in their research application.
- Most particle physics conferences now have an education and outreach session.

The education and outreach project of ATLAS has described how the ATLAS experiment has explored, and still is exploring the inner structure of matter and some of the mysteries we still encounter. Our ambition was to make the images and stories understandable and available to everybody, sometimes more detailed and challenging for teachers and students, but always aesthetically interesting.

Students and the general public are excited and motivated by learning about the amazing developments in contemporary physics. Our goal of getting them to pursue these interests has been fulfilled.
We are happy to have been part of this fascinating endeavour together with so many of our ATLAS colleagues.

# Acknowledgements

We had the great support of Peter Jenni in particular, but also Torsten Åkesson, Dave Charlton, Fabiola Gianotti, Marzio Nessi, Markus Nordberg, and ATLAS management in general.

During the period of this report (1996-2012), the education and outreach program benefitted from the outstanding work and contributions of many people, including
Hans Peter Beck, Uta Bilow, Helfried Burckhart, Laurent Chevalier, Manuela Cirilli, Juliette Davenne, Kaushik De, Markus Elsing, Jean Ernwein, Pauline Gagnon, James Gillies, Steve Goldfarb, Neal Hartman, Michael Hauschild, Dirk Hoffmann, Vato Kartvelishvili, Michael Kobel, Christine Kourkoumelis, Josef Kristofoletti, Alison Lister, Amelia Maio, Fabienne Marcastel, Claudia Marcelloni, Katie McAlpine, Sascha Mehlhase, David Milstead, Alexander Oh, Joao Pequenao, Ceri Perkins, Connie Potter, Emma Sanders, Paul Schaffner, Reinhard Schwienhorst, Chris Thomas, Josiane Uwantege, Bob van Gijzel, Peter Watkins, Tiina Wickstroem, Monika Wielers. Katie Yurkewicz.   With apologies to those people we may have forgotten.  After 2012, the Education and Outreach program has had outstanding contributions from additional people.

--------  END -----------------------